\documentclass{article}

\usepackage[final,sglblindworkshop]{neurips_2026}
\workshoptitle{7th Muslims in ML (MusIML) Workshop}

\usepackage[utf8]{inputenc}
\usepackage[T1]{fontenc}
\usepackage{hyperref}
\hypersetup{pdfauthor={Ameen Ur Rehman, Ali Nasir},
  pdftitle={Federated Learning for Unmanned Aerial Systems: A Screened Bibliometric Census of an Emergent Sub-Literature},
  pdfsubject={Bibliometric census}, pdfkeywords={federated learning, UAV, bibliometrics, Scopus, OpenAlex}}
\usepackage{url}
\usepackage{booktabs}
\usepackage{amsfonts}
\usepackage{amsmath}
\usepackage{nicefrac}
\usepackage{microtype}
\usepackage{xcolor}
\usepackage{graphicx}
\usepackage{caption}
\usepackage{subcaption}
\usepackage{tikz}
\usetikzlibrary{arrows.meta,positioning,calc}
\usepackage{placeins}
\usepackage{enumitem}
\setlist{nosep}

\definecolor{stageblue}{HTML}{2E5A88}
\definecolor{stageteal}{HTML}{1E7A72}
\definecolor{stageamber}{HTML}{D49A1E}
\definecolor{stagered}{HTML}{B5473A}
\definecolor{stagegray}{HTML}{6B6F76}
\definecolor{stagegrayL}{HTML}{ECEDEF}

\title{Federated Learning for Unmanned Aerial Systems:\\A Screened Bibliometric Census of an Emergent Sub-Literature}

\author{%
  Ameen Ur Rehman \\
  Department of Control and \\
  Instrumentation Engineering \\
  King Fahd University of \\
  Petroleum and Minerals \\
  Dhahran 31261, Saudi Arabia \\
  \texttt{g202527430@kfupm.edu.sa} \\
  \And
  Ali Nasir \\
  Department of Control and \\
  Instrumentation Engineering \\
  King Fahd University of \\
  Petroleum and Minerals \\
  Dhahran 31261, Saudi Arabia \\
  \texttt{ali.nasir@kfupm.edu.sa} \\
}

\begin{document}

\maketitle

\begin{abstract}
We census the literature at the intersection of federated learning (FL) and unmanned aerial systems (UAS) with a broad Scopus query that combines FL, UAV, and delivery/logistics terms (Computer Science and Engineering, 2022--2026). The export holds \textbf{14,965 records}, but only \textbf{7.9\%} of the 14,945 unique records mention both an FL term and a UAV term in title, abstract, or author keywords. That screen leaves \textbf{1,185 records}: 90 in 2022, 361 in 2025, and 332 already indexed for 2026. The screened share of each year's raw output stays near 7.5\%, so the growth is not an artefact of the screen. Term-definition, document-type, and complete-year checks leave the leading venue (\emph{IEEE Internet of Things Journal}, 111 papers, $h{=}30$) and the leading country (China, 628 affiliation instances) unchanged. A manual read of 30 records finds 14 of 26 full papers (54\%, 95\% CI 35--71\%) substantively about FL for UAVs, so the corpus counts keyword co-occurrence, not only focused work. Counting authors by Scopus ID gives 4,095 authors with Lotka exponent $n{=}3.15$ ($R^2{=}0.996$); name strings would merge distinct people and understate the author count by 31\%. An OpenAlex export under the same screen yields 147 records, 74\% of them in the Scopus corpus, but the two exports are not matched populations, so we do not read the volume gap as coverage alone. Finally, of 228 records with a delivery-adjacent term, only 5 (2.2\%) mention \emph{parcel}, \emph{last-mile}, or \emph{courier} (14, or 6.1\%, if \emph{package} counts); the rest reflect transportation, traffic, and generic-logistics idiom. Keyword screening alone therefore cannot isolate last-mile drone-delivery research from the surrounding UAV-networking literature.
\end{abstract}

\section{Introduction}
\label{sec:intro}

Federated learning, which trains models across data that cannot be centralized \citep{mcmahan2017,kairouz2021}, and unmanned aerial systems, which are flying compute and communication endpoints with energy, connectivity, and trust constraints \citep{brik2020,qu2021}, developed separately and met only recently. The intersection is young enough that nobody has mapped it: who publishes it, how fast it grows, and whether it is large enough to support a narrower application claim such as drone-based last-mile delivery.

We attempt that map and, along the way, expose a methodological problem for any keyword-based scoping of this space. A Scopus query built the usual way, OR-ing synonyms inside each concept and AND-ing the three concepts, returns 14,965 records. Most are noise: ``distributed'' and ``deliver*'' sweep in distributed-systems and data-delivery papers unrelated to FL for UAVs. A raw hit count is not a corpus, so we screen before analyzing and report every step of the funnel (Figure~\ref{fig:funnel}), following the retrieval-to-corpus discipline of recent Scopus science mapping at a comparable workshop \citep{nadeem2026}.

We ask three questions of the screened corpus. How large is it and how fast does it grow? Where, by whom, and from which countries is it published? Can a delivery-specific sub-literature be isolated from it? The third matters most for anyone positioning work in FL for drone delivery rather than FL for UAV networks in general, and the answer is that, by keyword alone, it cannot.

\paragraph{Contributions.}
\begin{itemize}
  \item A screened, reproducible census of 1,185 FL-for-UAV records (2022--2026) in which the screening step is itself a result: 92\% of the deduplicated Boolean retrieval fails the topical test, and the leading venue, the leading country, and the top-ten lists are re-tested under seven alternative screens.
  \item Venue, authorship, and geographic structure of the sub-literature, with author identity resolved by Scopus ID and citation impact reported as both mean and median, cohort-adjusted.
  \item A cross-database check against OpenAlex that reports the overlap of the two screened corpora and states what the mismatched query execution can and cannot support.
  \item A decomposition showing that ``delivery'' is not a usable keyword filter here, with consequences for how narrower application claims should be scoped.
\end{itemize}

\section{Related work}
\label{sec:related}

\paragraph{FL and UAV networks.} The FL literature is well surveyed \citep{kairouz2021,li2020,yang2019,imteaj2022}. Within it, a smaller strand treats UAVs as FL participants or coordinators: \citet{brik2020} on use cases, challenges, and open problems, \citet{qu2021} on decentralized architectures for UAV networks, and \citet{alaya2025} on blockchain-FL integration in UAV applications. All three sit inside our corpus and illustrate its pattern: FL for UAVs is a systems and communications literature first, with learning embedded in a networking problem.

\begin{figure}[t]
  \centering
  \resizebox{\linewidth}{!}{%
  \begin{tikzpicture}[x=1cm,y=1cm,font=\sffamily\footnotesize]
    \tikzset{
      mb/.style={draw=stageblue,thick,rounded corners=2pt,fill=stageblue!7,minimum width=7.7cm,minimum height=1.05cm,anchor=north west},
      ex/.style={draw=stagegray,rounded corners=2pt,fill=stagegrayL,text width=5.2cm,anchor=west,align=left,inner sep=3pt,font=\sffamily\scriptsize},
      ar/.style={-{Stealth[length=2mm]},thick,draw=stageblue!80},
      dsh/.style={densely dashed,draw=stagegray,thick}
    }
    \def\pitch{1.56}
    \foreach \i/\t/\s/\c/\f in {
      0/{Raw Scopus export}/{FL $\wedge$ UAV $\wedge$ delivery terms; CS+Eng.; 2022--26}/{14,965}/1.0,
      1/{Deduplicated}/{by EID, then exact title}/{14,945}/0.9987,
      2/{FL-side term in title, abstract, keywords}/{\texttt{federat*} or \texttt{distribut*}: 39.6\% of 14,945}/{5,921}/0.3957,
      3/{Screened corpus (FL $\wedge$ UAV)}/{adds \texttt{uav}, \texttt{drone}, \ldots: 7.9\% of 14,945}/{1,185}/0.0792,
      4/{Delivery-adjacent term}/{135 with core terms; 228 adding logistics, transportation}/{228}/0.0152} {
      \node[mb] (b\i) at (0,-\i*\pitch) {};
      \node[anchor=north west,align=left] at ($(b\i.north west)+(0.15,-0.08)$) {\textbf{\t}\\[-2pt]{\scriptsize\color{stagegray}\s}};
      \node[anchor=north east] at ($(b\i.north east)+(-0.15,-0.06)$) {\large\bfseries\c};
      \fill[stagegrayL] ($(b\i.south west)+(0.15,0.1)$) rectangle ($(b\i.south east)+(-0.15,0.17)$);
      \fill[stageblue] ($(b\i.south west)+(0.15,0.1)$) rectangle ($(b\i.south west)+(0.15+7.4*\f,0.17)$);
    }
    \draw[stageblue,line width=1.5pt,rounded corners=2pt] (b3.north west) rectangle (b3.south east);
    \foreach \a/\b in {0/1,1/2,2/3,3/4} {\draw[ar] (b\a.south) -- (b\b.north);}
    \foreach \i/\txt in {
      1/{$-$20 duplicate records},
      2/{$-$9,024 without an FL-side term},
      3/{$-$4,736 without a UAV term\\(only 32, or 0.7\%, contain a broader UAV term)},
      4/{$-$957 without a delivery term}} {
      \pgfmathsetmacro{\yy}{-\i*\pitch+0.26}
      \node[ex] (e\i) at (8.5,\yy) {\txt};
      \draw[dsh] (3.85,\yy) -- (e\i.west);
    }
    \node[ex,draw=stageteal,fill=stageteal!8,text width=7.3cm,anchor=north west] (rob) at (0,-4*\pitch-1.3)
      {\textbf{Robustness of the 1,185} (Section~\ref{sec:sensitivity}): narrow 740; FL-broadened 1,191; UAV-broadened 1,188; plural \texttt{UAVs} 1,215; Conference Reviews removed 1,126.};
    \node[ex,draw=stageamber,fill=stageamber!9,anchor=north west] (par) at (8.5,-4*\pitch-1.3)
      {\textbf{Parcel cue} (Section~\ref{sec:delivery}): 5 of 228 with \emph{parcel, last-mile, courier}; 14 if \emph{package} counts.};
  \end{tikzpicture}}
  \caption{From query to analytic corpus. Each box shows the record count; the bar under it is the share of the 14,965 raw records. Grey boxes list what each step removes. The teal box lists corpus sizes under the alternative screens tested in Section~\ref{sec:sensitivity}; the amber box gives the parcel-specific count for the delivery analysis in Section~\ref{sec:delivery}.}
  \label{fig:funnel}
\end{figure}
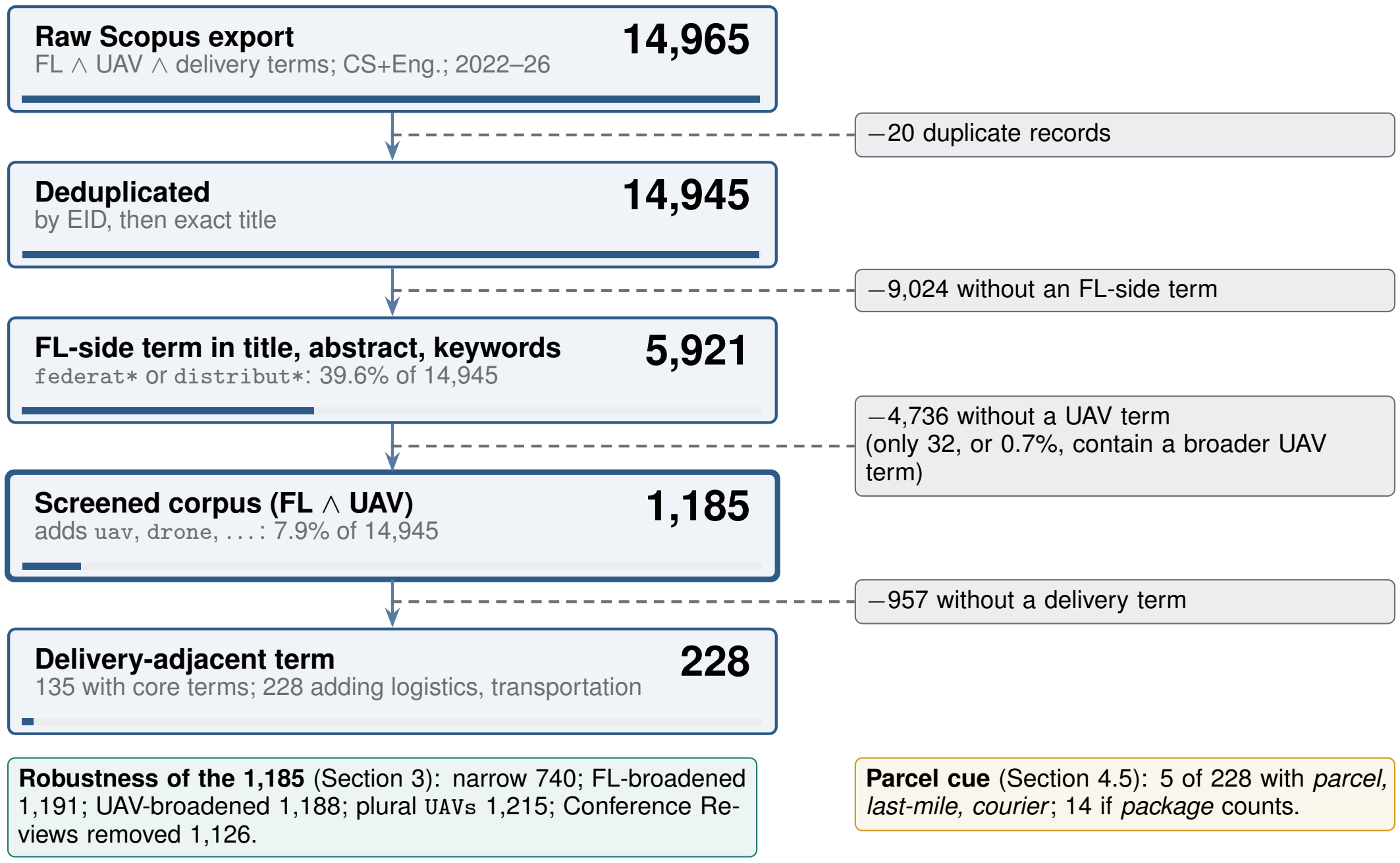

\paragraph{Bibliometric method and screening protocols.} We follow standard science-mapping practice \citep{zupic2015,donthu2021,aria2017} within the science of science \citep{fortunato2018,clauset2017}, and use Lotka's law \citep{lotka1926} with the $h$- and $g$-indices \citep{hirsch2005,egghe2006}. Systematic-review frameworks differ in detail but share staged reporting: PRISMA \citep{page2021} specifies identification, screening, and inclusion with manual adjudication of each record, and SALSA \citep{grant2009} frames a review as search, appraisal, synthesis, and analysis. We borrow the stage-wise reporting (Figure~\ref{fig:funnel}) but replace per-record adjudication with a regular-expression screen plus a sampled manual check. That trades topical precision for exact reproducibility and scale, and Section~\ref{sec:validation} measures the price. Previous FL bibliometrics map the field as a whole \citep{farooq2021,kumar2025}; we did not find a census restricted to its UAV intersection. The closest methodological precedent is a Scopus-based mapping paper at a sister workshop that moved from 594 raw records to a screened corpus of 445 \citep{nadeem2026}.

\paragraph{Database coverage.} Scopus, Web of Science, and open indexes differ in coverage and metadata quality \citep{mongeon2016,martinmartin2018,visser2021}. OpenAlex \citep{priem2022} is broad and in some respects a superset of Scopus, with metadata gaps that matter for abstract-based screens \citep{alperin2024}. We cross-check against OpenAlex but treat the result as a check on one pipeline, not a coverage estimate.

\section{Data and method}
\label{sec:method}

\paragraph{Query and funnel.} We queried Scopus for \texttt{ALL(("federated learning" OR "federated machine learning" OR "distributed learning") AND (drone* OR uav* OR "unmanned aerial vehicle*" OR uas* OR quadcopter* OR multicopter*) AND (deliver* OR logistics OR transportation OR routing OR scheduling OR "last-mile delivery" OR parcel* OR package* OR fleet*)) AND PUBYEAR > 2021 AND PUBYEAR < 2027}, restricted to the Computer Science and Engineering subject areas. The export was retrieved in July 2026. Figure~\ref{fig:funnel} gives the three steps from query to corpus.
\begin{enumerate}
  \item \textbf{Raw export: 14,965 records}, the row count of the exported CSV (Scopus's 45-field schema).
  \item \textbf{Deduplication: 14,945 records.} Deduplicating on EID, then exact title, removes 20 records. This set is the denominator for every ``\% of the corpus'' figure.
  \item \textbf{Topical screening: 1,185 records.} A record enters the corpus if its title, abstract, or author keywords match, case-insensitively, \emph{both} an FL term (\texttt{federat*} or \texttt{distribut*}) \emph{and} a UAV term (\texttt{\textbackslash buav\textbackslash b}, \texttt{drone}, \texttt{unmanned aerial vehicle}, \texttt{unmanned aircraft system}, \texttt{\textbackslash buas\textbackslash b}). Of 14,945 records, 5,921 (39.6\%) pass the FL test; of those, 1,185 (20.0\%) also pass the UAV test. The joint rate is 7.9\%; the 39.6\% describes the FL filter alone.
\end{enumerate}
We do not require a delivery term to survive screening: doing so would turn a terminology confound into an apparent finding (Section~\ref{sec:delivery}). The query strings (including the OpenAlex string and export dates), regular expressions, screened-EID list with document types, country list, and the scripts that recompute every number are released with the paper; Scopus licensing prevents redistributing the records themselves, which can be re-exported from the stated query. Delivery patterns are given in Section~\ref{sec:delivery}.

\paragraph{Sensitivity of the term definitions.}
\label{sec:sensitivity}
Table~\ref{tab:sensitivity} re-runs the screen under seven alternative definitions on the same 14,945 records. The \emph{narrow} screen keeps \texttt{federat*} only, dropping records about distributed-but-not-federated training. The \emph{FL-broadened} screen adds split, vertical, personalized FL and collaborative learning. The \emph{UAV-broadened} screen adds quadrotor, aerial robot, and MAV. The \emph{plural} screen accepts \texttt{UAVs} written without the singular form, which the reported pattern misses; it adds 30 records (2.5\%). We also drop the 59 Conference Review records, restrict to the four complete years, and apply a \emph{strict core} that needs \texttt{federat*} and a UAV term in the title or author keywords. In every row \emph{IEEE Internet of Things Journal} leads the venues and China leads the countries. The reported top-ten venues keep at least 7 of their places in every screen, and the top-ten country lists are identical in all of them. We did not test the delivery-side term list beyond the final classification, and we cannot test the subject-area filter without re-querying Scopus. Under the narrow screen the top five venues are the same five titles (69, 26, 26, 19, and 15 papers), China again leads the countries (337 instances), and annual output stays monotone through 2025 (58, 111, 161, 222).

\begin{table}[t]
  \caption{Sensitivity of the screened corpus to the term definition, on the same 14,945 deduplicated records. Corpus (\%) is the share of the 14,945 records. ``Overlap'' is how many of the reported screen's top-10 venues remain in the top 10; the top-10 countries are identical in every row.}
  \label{tab:sensitivity}
  \centering
  \footnotesize
  \setlength{\tabcolsep}{4pt}
  \begin{tabular}{lrrrrr}
    \toprule
    Screen & Corpus (\%) & Top venue & Top country & Overlap \\
    \midrule
    Narrow (\texttt{federat*} only)          &   740 (5.0) &  69 &  337 &  8/10 \\
    \textbf{Reported}                         & \textbf{1,185 (7.9)} & \textbf{111} & \textbf{628} & n/a \\
    FL-broadened                              & 1,191 (8.0) & 112 &  632 & 10/10 \\
    UAV-broadened                             & 1,188 (7.9) & 111 &  630 & 10/10 \\
    Plural \texttt{UAVs} accepted             & 1,215 (8.1) & 112 &  635 & 10/10 \\
    Conference Reviews removed                & 1,126 (7.5) & 111 &  628 & 10/10 \\
    Complete years only (2022--25)            &   853 (5.7) &  83 &  439 &  7/10 \\
    Strict core (title/keywords)              &   466 (3.1) &  50 &  247 &  9/10 \\
    \bottomrule
  \end{tabular}
  \par\smallskip\raggedright\scriptsize ``Top venue'' counts papers in \emph{IEEE Internet of Things Journal}; ``top country'' counts affiliation instances for China.
\end{table}

\paragraph{Validation of the screen.}
\label{sec:validation}
Stable corpus size under different definitions does not show that matches are about FL for UAVs. We read the title and abstract of a random sample of 30 of the 1,185 records. Four (13.3\%; 95\% Wilson interval 5.3--29.7\%) are Scopus ``Conference Review'' records, proceedings front matter whose table-of-contents abstract matches two unrelated chapters; 59 records (5.0\% of the corpus) share this document type and are mechanically identifiable. Of the 26 remaining papers, 14 (53.8\%; Wilson 95\% interval 35.5--71.2\%) substantively apply FL within a UAV system, and 12 matched on co-occurrence alone, most often \texttt{distribut*} in a non-learning sense (``distributed denial-of-service'') or FL named in passing. Applied to the 1,126 non-Conference-Review records, the central estimate is about 600 focused papers (interval 400--800). We report this because it is what we found: 1,185 counts co-occurrence, and on this sample about half of it studies FL for UAVs. The sample is small and was read by one rater (Section~\ref{sec:limits}). A separate automated check on the screen's other error type supports the UAV side: of the 4,736 records that pass the FL test but fail the UAV test, 32 (0.7\%) contain a broader UAV term, so the UAV test discards few true positives. A hand check of twenty rejected records found each a genuine false positive.

Do the rankings survive restriction to the better-matched records? Thirty records are too few to re-rank directly, so we partition the corpus by the term that drove the match. The 740 records matching \texttt{federat*} (62\% of the corpus) and the 445 matching only \texttt{distribut*}, the pattern behind most sample false positives, each lead on \emph{IEEE Internet of Things Journal} (69 and 42 papers) and China (337 and 291 instances). Removing the Conference Reviews leaves both the top-five venues and the top-five countries in the same order. Residual noise inside either partition could still move positions lower in the list; settling that would need manual screening of all 1,185 records.

\paragraph{Measures.} We use the measures of prior large-scale bibliometrics. Lotka's law is fit by least squares in log-log space over $x\le10$ papers per author, with authors identified by Scopus Author ID and, for comparison, by name string. We compute $h$- and $g$-indices per venue, and country affiliation by lookup of the trailing token of each affiliation string. For citation impact we report mean and median citations per paper, and re-compute both on 2022--2024 papers only because citations depend on paper age. Year trends stop at 2025: 2026 was incomplete at retrieval, and its 332 records stay in aggregate totals but are drawn as a partial year.

\section{Results}
\label{sec:results}

\subsection{The raw retrieval overstates the field by an order of magnitude}
\label{sec:general}

\begin{figure}[t]
  \centering
  \begin{subfigure}[b]{0.43\textwidth}\includegraphics[width=\linewidth]{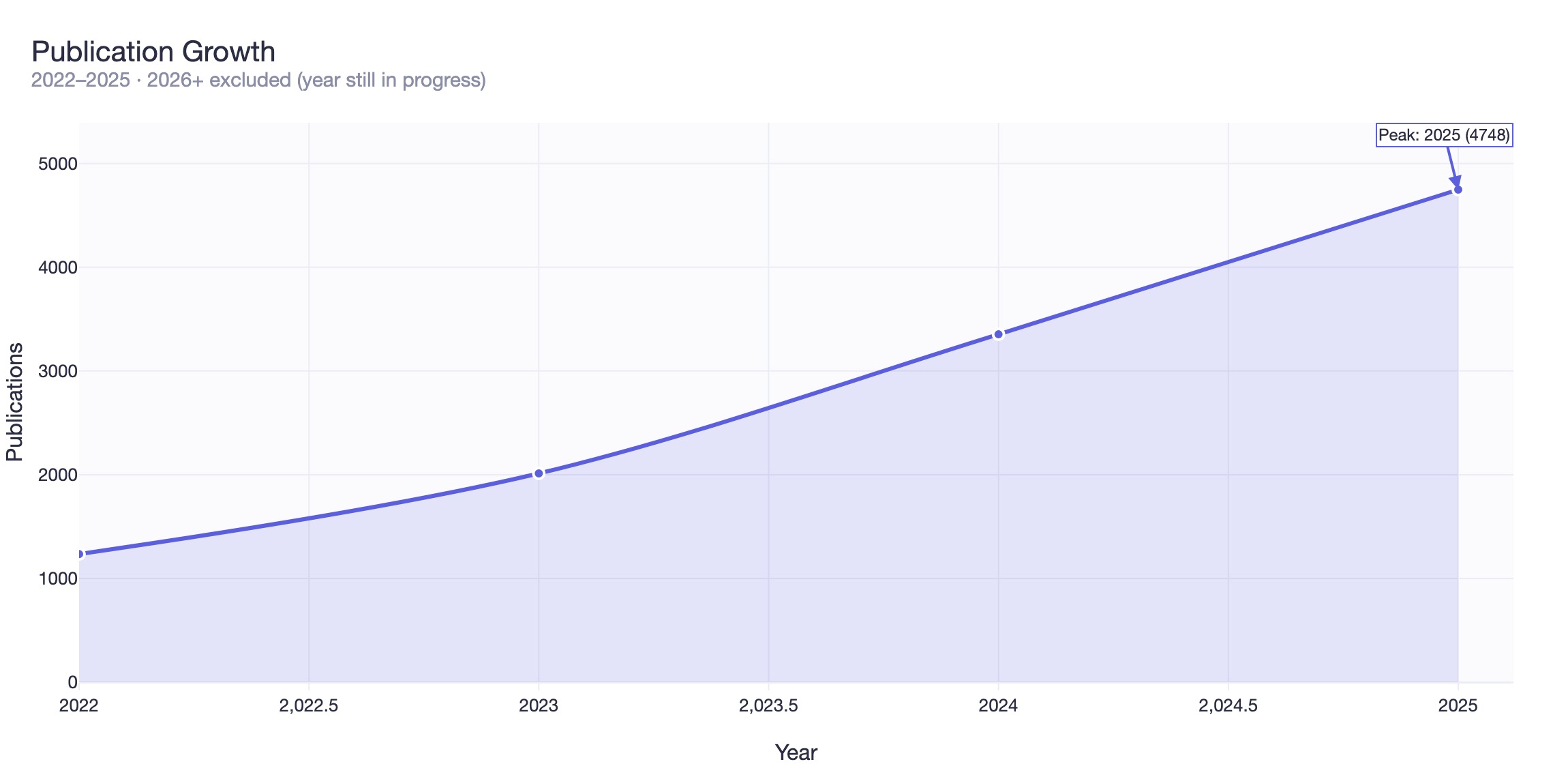}\caption{Raw retrieval, annual output.}\end{subfigure}\hfill
  \begin{subfigure}[b]{0.44\textwidth}\includegraphics[width=\linewidth]{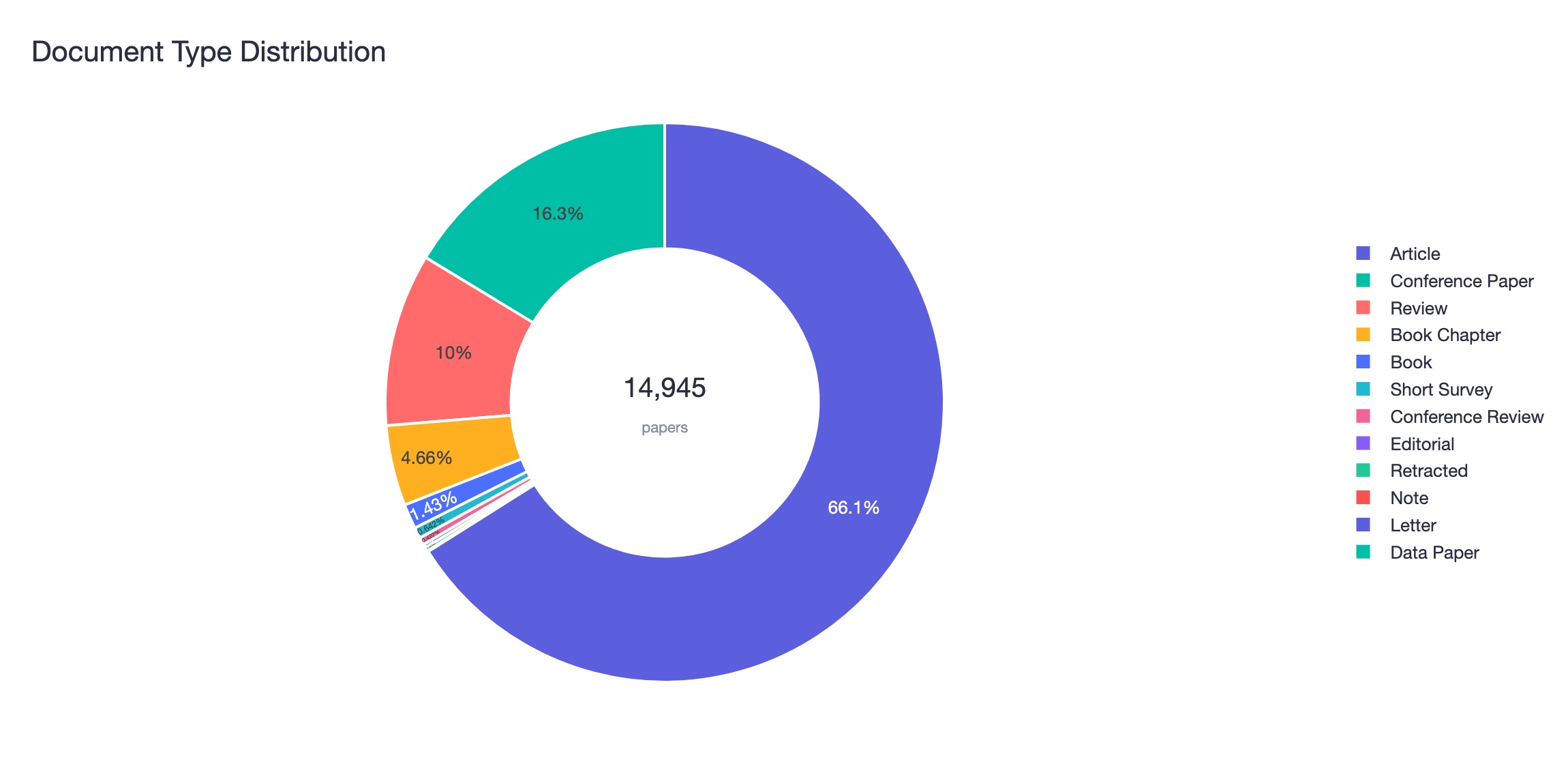}\caption{Raw retrieval, document type.}\end{subfigure}
  \caption{The general landscape of the raw retrieval, before topical screening. (a) Annual output already shows the same accelerating shape later confirmed for the screened corpus (Figure~\ref{fig:growth-venues}a). (b) The raw pool is journal-dominated and roughly one-third open access; neither pattern requires a record to be about FL for UAVs, which is why screening is needed before either figure can be attributed to this literature.}
  \label{fig:general-landscape}
\end{figure}

The deduplicated pool of 14,945 records is 12.6 times the corpus. It grew from 1,235 records in 2022 to 4,748 in 2025 (Table~\ref{tab:annual}, Figure~\ref{fig:general-landscape}a), is journal-dominated (\emph{Article} 66.1\%, \emph{Conference Paper} 16.3\%, \emph{Review} 10.0\%; Figure~\ref{fig:general-landscape}b), spans 2,770 venues and 23,783 name strings, and is 33.1\% open access. It describes everything the query swept in, on-topic or not. Mean citations per record fall from 56.5 (2022) to 8.1 (2025) because later cohorts have had less time to accumulate citations.

The corpus tracks the pool closely. Its share of each year's raw output is 7.3, 7.8, 7.3, and 7.6\% for 2022--2025 (Table~\ref{tab:annual}), so the screen does not manufacture a trend. The corpus itself grows from 90 to 157, 245, and 361 papers (Figure~\ref{fig:growth-venues}a; 59\% per year compounded over 2022--2025, against 57\% for the raw pool). The 332 records already indexed for roughly seven months of 2026 approach the full-year 2025 total, consistent with continued acceleration. The screened corpus is also more conference-heavy than the pool (conference papers 22.3\% against 16.3\%) and is 27.3\% open access.

\begin{table}[t]
  \caption{Annual production and citation accumulation for the raw retrieval (14,945 records) against the screened corpus (1,185). 2026 is a partial year (retrieved in July). Means decline with cohort age.}
  \label{tab:annual}
  \centering
  \small
  \begin{tabular}{lrrrrr}
    \toprule
    Year & Raw records & Total citations & Mean cit./record & Screened & Screened share \\
    \midrule
    2022 & 1,235 & 69,780 & 56.50 &  90 & 7.3\% \\
    2023 & 2,012 & 79,126 & 39.33 & 157 & 7.8\% \\
    2024 & 3,353 & 71,227 & 21.24 & 245 & 7.3\% \\
    2025 & 4,748 & 38,415 &  8.09 & 361 & 7.6\% \\
    2026* & 3,597 &  3,818 &  1.06 & 332 & 9.2\% \\
    \bottomrule
  \end{tabular}
\end{table}

\subsection{Venues and geography}
\label{sec:venues}

\begin{figure}[t]
  \centering
  \begin{subfigure}[b]{0.43\textwidth}\includegraphics[width=\linewidth]{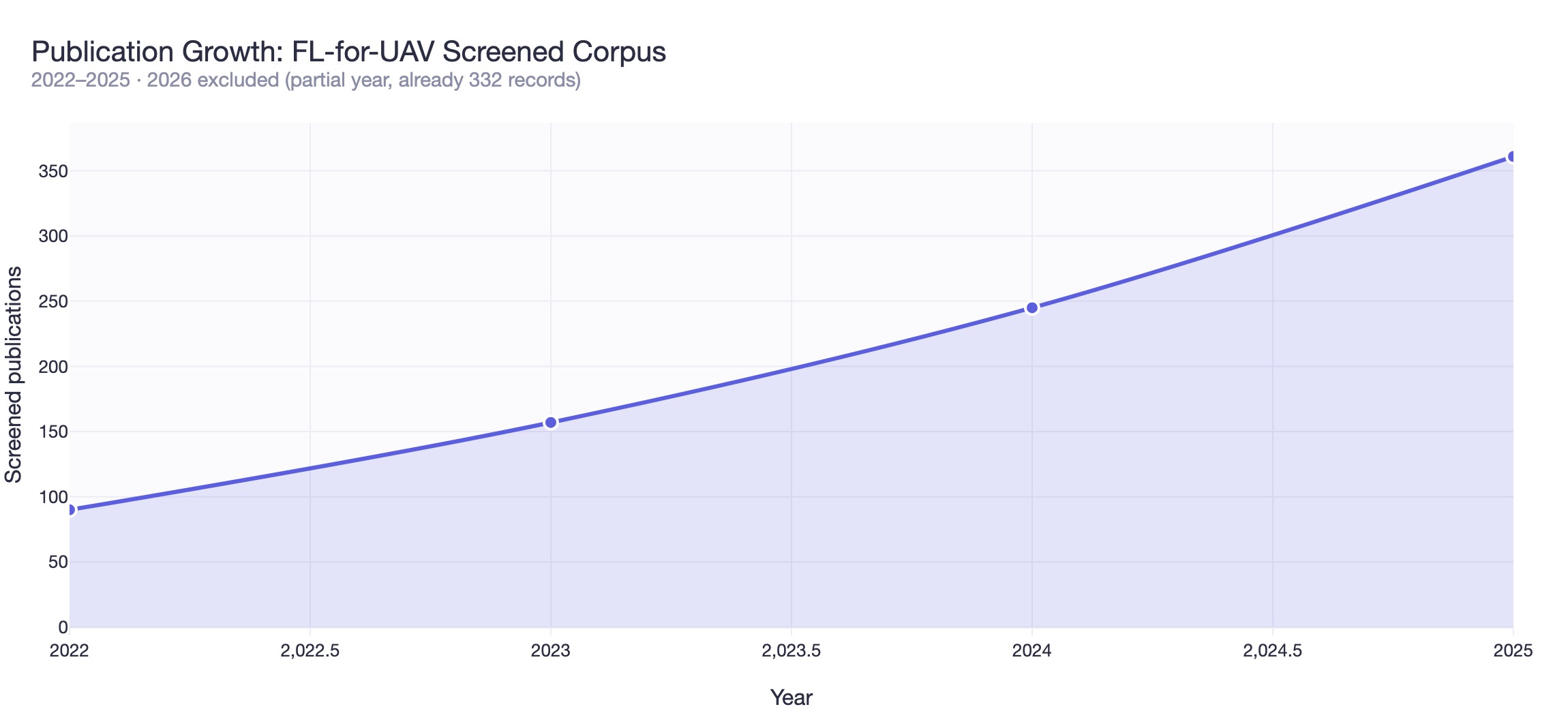}\caption{Annual output.}\end{subfigure}\hfill
  \begin{subfigure}[b]{0.44\textwidth}\includegraphics[width=\linewidth]{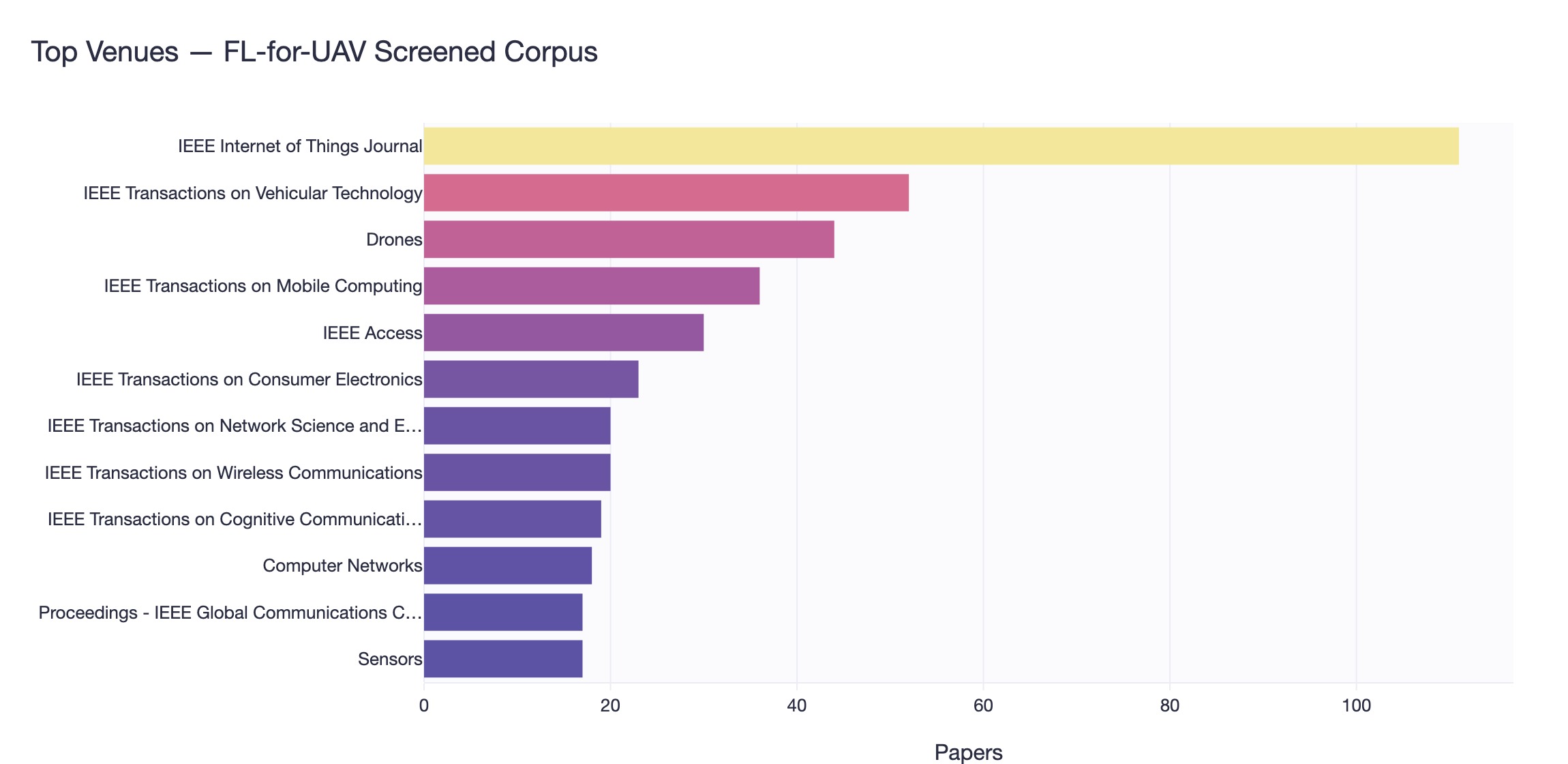}\caption{Top venues by volume.}\end{subfigure}
  \caption{Growth and venue concentration in the screened FL-for-UAV corpus. (a) Annual output is monotone and accelerating through the last complete year (2025); the partial 2026 count is excluded from the curve to avoid reading an in-progress year as a decline. (b) Venue ranking by volume.}
  \label{fig:growth-venues}
\end{figure}

Publication concentrates in IEEE networking and systems venues (Figure~\ref{fig:growth-venues}b). \emph{IEEE Internet of Things Journal} leads with 111 papers ($h{=}30$, $g{=}52$), followed by \emph{IEEE Transactions on Vehicular Technology} (52) and \emph{Drones} (44), the only non-IEEE title in the top five. The top venue holds 9.4\% of the corpus, the top five 23.0\%, and the top ten 31.5\%; the corpus spans 434 venues, 303 of which appear once, and about half of all records (50.5\%) are in venues with ``IEEE'' in the title. The most-cited item is a survey of FL for resource-constrained IoT devices \citep{imteaj2022} with 739 citations at retrieval; the next two (660 and 624) are broad 6G and non-terrestrial-network surveys that mention FL and UAVs among many enabling technologies, a pattern that recurs below.

\begin{figure}[t]
  \centering
  \includegraphics[width=0.48\linewidth]{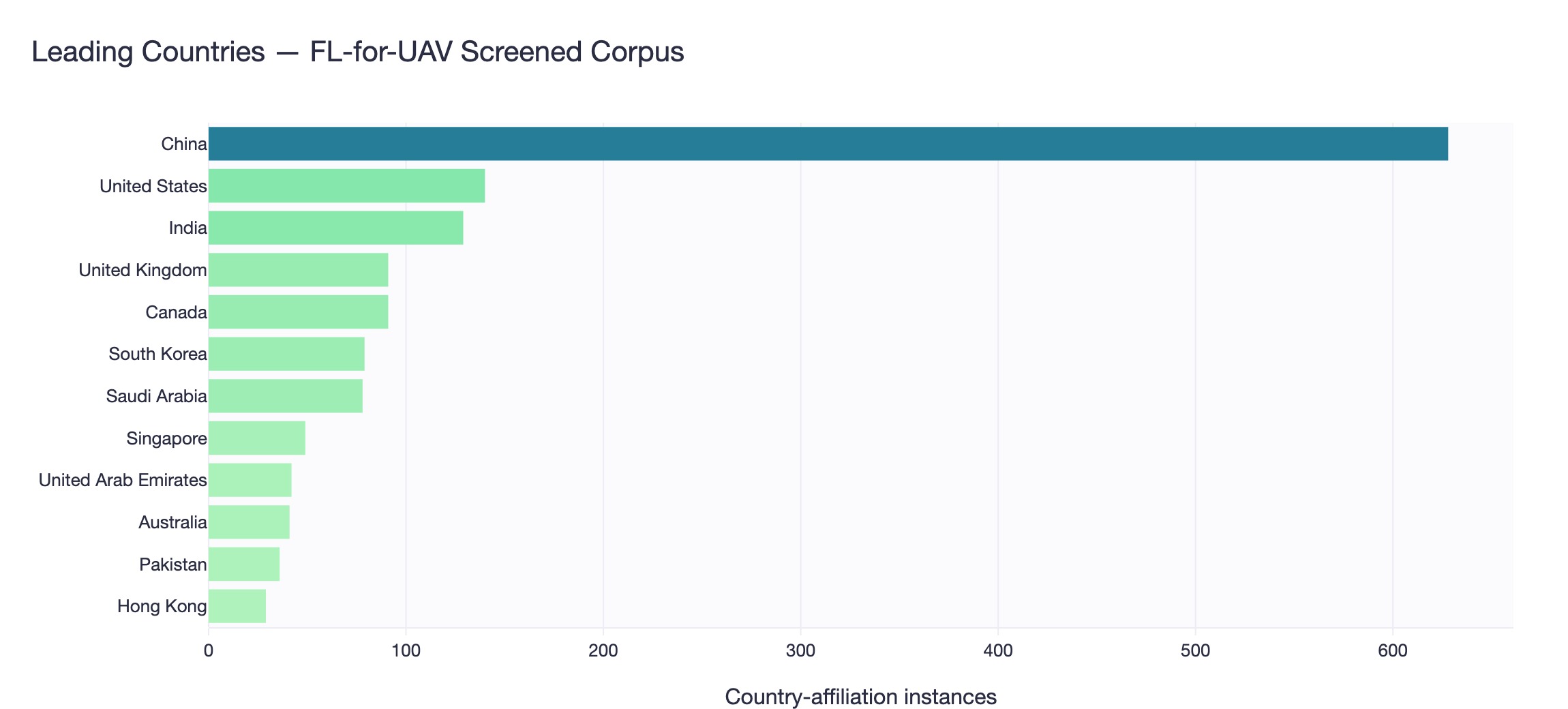}
  \caption{Leading countries by country-affiliation instances in the screened corpus.}
  \label{fig:countries}
\end{figure}

China leads by volume with 628 affiliation instances, 4.5 times the United States (140) and ahead of India (129) (Figure~\ref{fig:countries}). Volume and citation impact diverge. Among the 19 countries with at least 15 papers, Germany (56.6 mean citations per paper, 17 papers), Australia (55.2, 41), and Hong Kong (31.2, 29) have the highest means. Means mislead in a heavy-tailed distribution, so we also report medians: the same three countries lead on the median (Hong Kong 21, Australia 20, Germany 12), and China's mean of 13.1 corresponds to a median of 3. Much of the gap is cohort age. 62\% of China-affiliated records are from 2025--2026, against 18\% for Germany. Restricted to 2022--2024 papers, the median is 16.5 for China, 14 for the United States, 15.5 for Germany, and 22 for Australia, so China's lower average is largely a cohort effect rather than lower per-paper impact. Germany's mean rests on only 17 papers.

\paragraph{Participation from Muslim-majority countries.} Because this workshop concerns Muslim communities in ML, we also count participation from Muslim-majority countries, using an explicit country list released with the code. They supply 16.9\% of the corpus's country-affiliation instances (312 of 1,848) and 215 of the 1,185 papers (18.1\%) have at least one such affiliation. Saudi Arabia is seventh in the corpus by volume (78 instances) and the United Arab Emirates ninth (42), followed by Pakistan (36) and Malaysia (26); the shares in the raw pool are similar (17.5\%, 4,434 of 25,394), led by Saudi Arabia (952) and the UAE (492). This is descriptive context, and the country list is approximate.

\subsection{Authorship and vocabulary}
\label{sec:authors}

\begin{figure}[t]
  \centering
  \begin{subfigure}[b]{0.43\textwidth}\includegraphics[width=\linewidth]{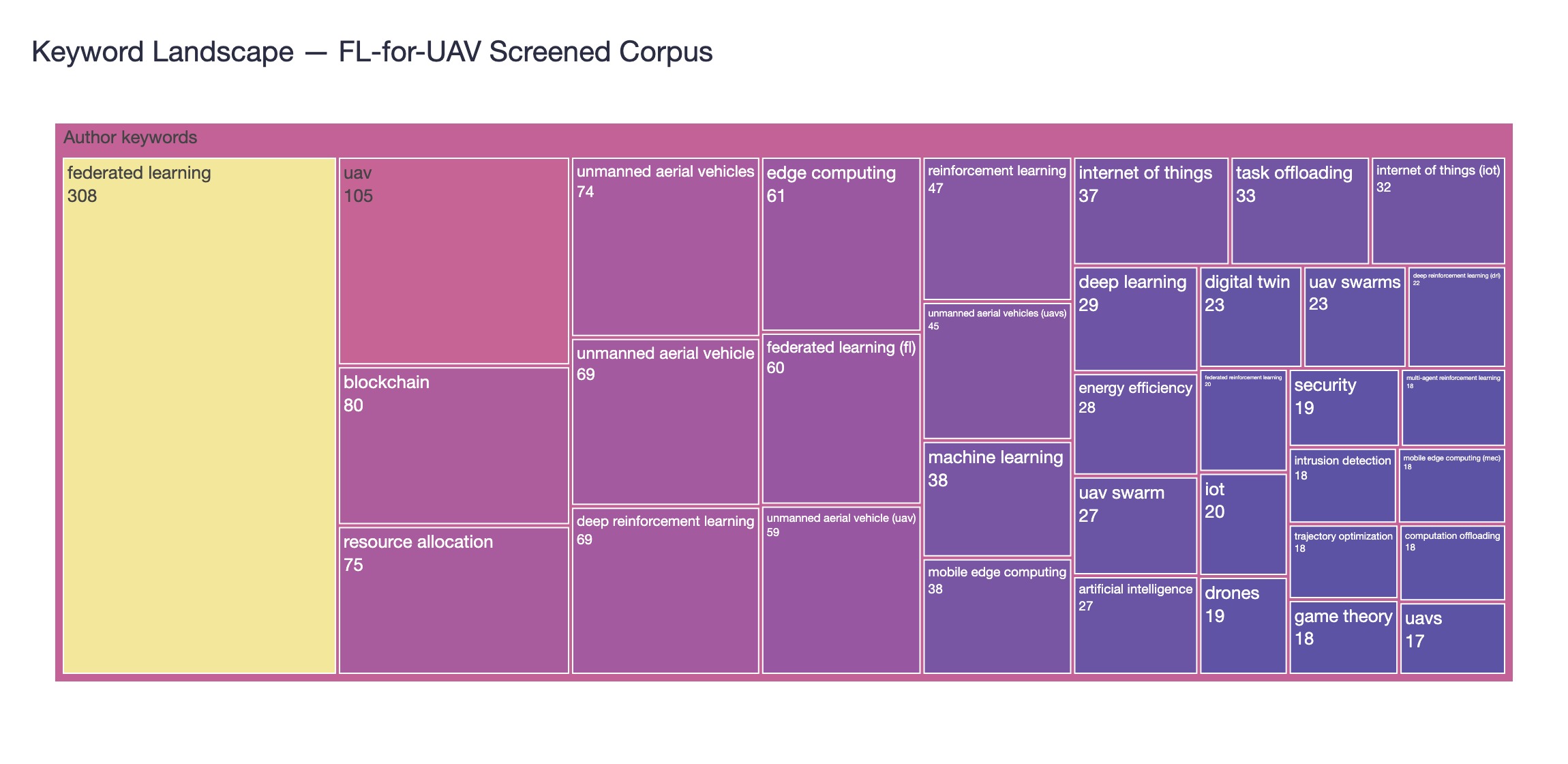}\caption{Top 35 author keywords.}\end{subfigure}\hfill
  \begin{subfigure}[b]{0.44\textwidth}\includegraphics[width=\linewidth]{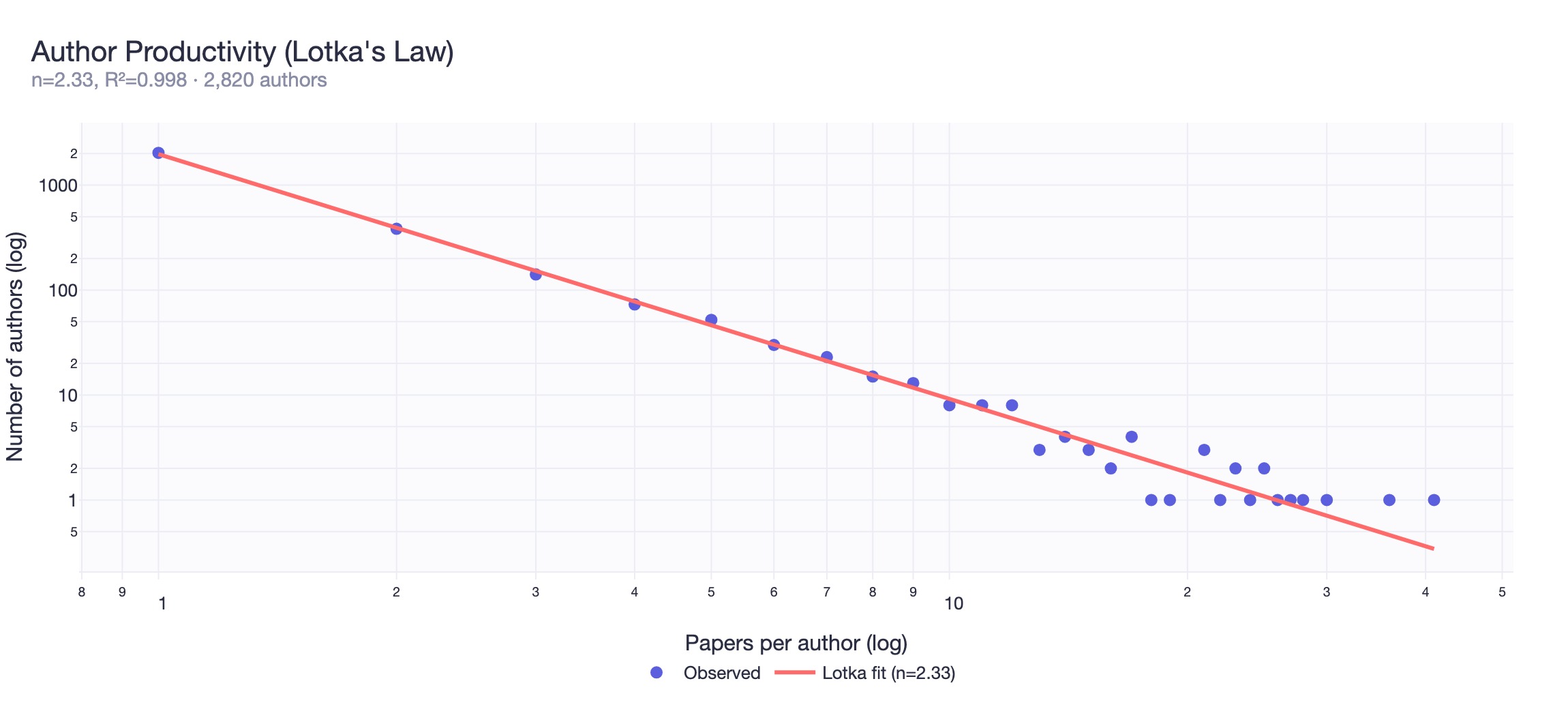}\caption{Author productivity (Lotka's law), by name string.}\end{subfigure}
  \caption{Vocabulary and authorship structure of the screened corpus. (a) Keyword frequency by author-supplied tag; variants such as \emph{uav} and \emph{unmanned aerial vehicles} are counted separately. (b) Lotka fit with authors identified by name string ($n{=}2.33$); fitting by Scopus Author ID gives $n{=}3.15$ with 4,095 authors (Section~\ref{sec:authors}).}
  \label{fig:keywords-lotka}
\end{figure}

Author identity changes the answer. By name string the corpus has 2,820 authors and a Lotka exponent of 2.33 ($R^2{=}0.998$), the fit drawn in Figure~\ref{fig:keywords-lotka}b. Scopus Author IDs give 4,095 authors and $n{=}3.15$ ($R^2{=}0.996$). The name strings merge people: 456 of the 2,820 strings (16.2\%) map to two or more IDs, one to as many as 24, and the most prolific ``author'' by name has 41 papers against 18 by ID. We therefore report the ID-based figures. An exponent above the classical $n{=}2$ \citep{lotka1926} means productivity decays faster than Lotka's baseline: most authors appear once and few appear often, as in a field still recruiting new entrants. The fit is also stable under the narrow screen ($n{=}3.26$, 2,459 authors). Mean team size is 4.46 authors.

Vocabulary is dominated by the UAV side. Grouping author-keyword variants into families (so that \emph{unmanned aerial vehicle}, \emph{UAVs}, and \emph{uav} count as one), UAV terms appear in 73\% of the 1,067 records that carry author keywords, FL terms in 48\%, reinforcement learning in 24\%, and resource-allocation, trajectory, or offloading terms in 21\% (Figure~\ref{fig:keywords-lotka}a shows the raw tags, with variants counted separately). Edge computing, security and privacy, IoT, and blockchain follow at 17\%, 16\%, 12\%, and 9\%. The learning component is carried by a systems vocabulary, the signature we also find for delivery.

\subsection{Cross-database check}
\label{sec:cross}

We ran a translated version of the three-concept query in OpenAlex, which has no equivalent of Scopus's \texttt{ALL()}, \texttt{PUBYEAR}, or \texttt{LIMIT-TO} operators. The two exports are not matched populations: Scopus returned 14,965 records and OpenAlex 222, so OpenAlex's retention under the screen (66.2\% against 7.9\%) mostly reflects how each platform executed the query, not a precision gap between databases. We therefore compare the two only after the same screen (Table~\ref{tab:cross}). The reported screen retains 147 OpenAlex records against 1,185 in Scopus (a ratio of 8.1); under the narrow screen the ratio is 5.2. Only 41.9\% of OpenAlex export rows carry an abstract, against 100\% of Scopus records, which depresses OpenAlex's abstract-based matches.

The overlap is more informative than the ratio. Matching on DOI, then normalized title, 110 of the 147 screened OpenAlex records appear in the Scopus export and 109 in the Scopus screened corpus. Taking OpenAlex as an independent reference set, the Scopus pipeline therefore recovers 74\% of it. The 37 OpenAlex-only records include 10 preprints and 20 articles and are not in Scopus at all; they reflect indexing differences, not screening. Conversely, the OpenAlex export reaches only 9.2\% of the Scopus corpus. The two databases converge on a similarly structured literature at different depths, which is partial external validation, not proof of completeness.

\begin{table}[t]
  \caption{Scopus and OpenAlex under the same screen. The raw exports are not matched populations, so retention is not comparable across rows. 109 of 147 OpenAlex records (74.1\%) are in the Scopus corpus.}
  \label{tab:cross}
  \centering
  \small
  \begin{tabular}{lrrrr}
    \toprule
    Database & Raw & Reported screen & Narrow screen & Retention \\
    \midrule
    Scopus   & 14,965 & 1,185 & 740 & 7.9\% \\
    OpenAlex &    222 &   147 & 143 & 66.2\% \\
    \bottomrule
  \end{tabular}
\end{table}

\subsection{Why ``delivery'' does not isolate a delivery-specific sub-literature}
\label{sec:delivery}

This is the finding most relevant to scoping FL for drone delivery as opposed to UAV networks in general. Restricting the corpus to records with a delivery-adjacent term (\emph{deliver}, \emph{last-mile}, \emph{courier}, \emph{parcel}, \emph{package}) yields 135 records; adding \emph{logistics} and \emph{transportation} yields 228, 20 of them Conference Reviews. We sort the 228 by the first matching rule, tested in this order: parcel, last-mile, or courier; \emph{package}; transportation or traffic; logistics; data, packet, or message delivery; any other \emph{deliver*} use. Parcel and package cues are tested first, so their counts are upper bounds on parcel-logistics content. Only \textbf{5 (2.2\%)} contain a parcel-specific term (\emph{parcel}, \emph{last-mile}, \emph{courier}); counting \emph{package} raises this to 14 (6.1\%); the reviewed version reported 13 (5.7\%) under a package rule we could not reproduce exactly, and the two definitions here bracket it. The rest match transportation or traffic idiom (41.2\%), other uses of \emph{deliver} (26.8\%), generic logistics (16.7\%), or data and packet delivery (9.2\%), with \emph{package} alone accounting for another 3.9\%. Non-parcel matches outnumber parcel matches between 15 and 45 to 1, depending on the definition. The three most-cited records among the 228 (660, 145, and 106 citations) are a 6G vehicle-to-everything survey, a UAV-assisted RIS survey, and a UAV-blockchain-FL taxonomy, none about parcel delivery. Isolating last-mile FL-for-UAV work would need manual or full-text screening; it does not form a keyword-addressable cluster.

\section{Discussion}
\label{sec:discussion}

\paragraph{The raw retrieval was not the corpus.} The largest number in this study is 14,965, not 1,185, and the gap between them is the main methodological result. A query built to maximize recall across three concept groups returns a pool that is 92\% off-topic by its own criteria, while the corpus's leading venue and country, and its top-ten lists, are stable across seven alternative screens. Anyone using a Boolean hit count as a field size should screen, report the funnel, and show that conclusions survive reasonable term changes.

\paragraph{Screening trade-offs and a hybrid workflow.} The regular-expression screen is reproducible and cheap but imprecise: about half of what it keeps is co-occurrence. A fully manual PRISMA-style screen of 14,945 records is out of reach for most groups, so a middle path follows: an automated screen to a few thousand candidates, document-type filters applied up front (Conference Reviews are 5.0\% of the corpus and 8.8\% of the delivery subset), a supervised relevance classifier trained on a double-annotated sample, and sampled manual adjudication to estimate precision with an interval. We tested the first two ingredients here and the last only on 30 records.

\paragraph{The delivery niche is a construction project, not yet a literature.} Work scoped to FL for drone-based last-mile delivery is positioned against a systems-engineering literature (FL for UAV networks, $N{=}1{,}185$), not an established delivery sub-literature, because the application has not yet accumulated distinct terminology. That is not evidence against the topic; early work in it competes with the UAV-networking mainstream, since no delivery-specific peer group is bibliometrically visible.

\section{Limitations and societal impact}
\label{sec:limits}

\paragraph{Limitations.} \textbf{(1) Screening precision is bounded.} Manual inspection finds about 54\% of the 26 sampled papers substantively about FL for UAVs, with a wide interval (35--71\%); the sample is small, read by one rater, and reports no inter-rater agreement. 1,185 is a keyword-matched count. A study of at least 100 records with two annotators and reported agreement would tighten this; the released code draws such a sample. \textbf{(2) Query scope.} Term-list sensitivity is tested on both the FL and UAV sides, but not on the delivery-side list or on the query's subject-area filter, which would need new retrievals. The UAV pattern also misses 30 records that use only the plural \emph{UAVs}. \textbf{(3) Affiliation parsing} is a lookup of the trailing token. A full ISO country list raises the share of papers with a resolved country from 89.7\% to 94.9\% and leaves the top-ten countries unchanged, but country totals remain lower bounds. \textbf{(4) Citation counts} depend on paper age and a few highly cited items, which is why we report medians and cohort-restricted values. \textbf{(5) Cross-database check} uses one alternative whose export is not a matched population, and OpenAlex rows often lack abstracts.

\paragraph{Societal impact.} Raw database hit counts are not corpus sizes, and treating them as such in a grant proposal, systematic review, or thesis proposal can overstate a field's maturity by more than an order of magnitude, as here (14,965 against 1,185). This follows from how multi-concept Boolean queries behave when a concept group contains a term with a common non-technical meaning (``distributed'', ``deliver''), and is not specific to this literature.

\section{Conclusion}

Federated learning for unmanned aerial systems is a real and fast-growing literature: 1,185 keyword-matched records since 2022, about half substantively on FL for UAVs on manual inspection, structured like a compressed version of the broader FL field. It is not a federated-learning-for-drone-delivery literature: the terms that appear to identify one instead identify a communications-engineering vocabulary, and only 2.2\% of the delivery-adjacent records carry a strict parcel cue. Both findings replace an assumed size and an assumed niche with measured ones, which a scoping decision needs and a raw hit count cannot give.

\begin{ack}
The authors thank the open-access scientometrics community whose methodological work underpins this analysis. No specific funding was received for this work, and the authors declare no competing interests.
\end{ack}

\FloatBarrier

\end{document}